\documentclass[aps,prb,twocolumn,groupedaddress,showpacs,amsmath,amssymb,amsfonts]{revtex4}

\usepackage{latexsym,amssymb}
\usepackage{graphicx,color,pstricks}
\usepackage{epsfig,epsf,bm}

\definecolor{gray}{rgb}{0.7,0.7,0.7}

\begin{document}

\title{Electrical control and interaction effects of the RKKY interaction in helical liquids}

\author{Yu-Wen Lee}
\email{ywlee@thu.edu.tw} \affiliation{Department of Applied Physics, Tunghai University, Taichung, Taiwan, R.O.C.}

\author{Yu-Li Lee}
\email{yllee@cc.ncue.edu.tw} \affiliation{Department of Physics, National Changhua University of Education,
Changhua, Taiwan, R.O.C.}

\date{\today}

\begin{abstract}
We study the RKKY interaction mediated by the helical edge states of a quantum spin Hall insulator in the presence
of the Rashba spin-orbital coupling induced by an external electric field and the electron-electron interaction.
We show that in the presence of the Rashba coupling, the RKKY interaction induced by the helical edge states
contains not only the Heisenberg-like and the Dzyaloshinskii-Moria terms but also the nematic-type term that is
not present originally, with the range functions depending on the strength of the Rashba coupling. We also show
that the electron-electron interaction changes the strength of the RKKY interaction by modifying the power of the
$1/|x|$ dependence of the range functions. In particular, by varying the strength of the interaction or the Rashba
coupling, there is an (impurity) quantum phase transition involving the sign change of the RKKY interaction at the
value of the Luttinger liquid parameter $K=1/2$. Since both the strength of the Rashba coupling and the chemical
potential of the helical edge states are electrically controllable by external gate voltages, our results not only
shed light on the nature of magnetic impurity correlations in the edge of a two-dimensional topological insulator,
but also pave a way to manipulate the qubits in quantum computing.
\end{abstract}


\maketitle

\section{Introduction}

Over the past few years, the study of topological states of matters with or without topological order has attracted
a lot of attention in the condensed matter community. The exploration of these new states of matter not only
enriches the traditional condensed matter physics, but also has important implications in other branches of physics,
such as quantum information science and quantum computing.~\cite{Nayak} Among these topologically nontrivial states,
a special class of states, known as the symmetry-protected-topological (SPT) states, has the special property that
while the excitations in the bulk are gapped and trivial, they contain robust gapless edge excitations protected by
the symmetries of the system. A prototypical example of the SPT state is the two-dimensional quantum spin Hall
insulator (QSHI) which has a finite gap for bulk excitations and two counter-propagating gapless edge modes with
opposite spin polarizations.~\cite{Kane,Qi_RMP} This helical edge state is guaranteed to be stable against the weak
interaction and disorder due to the time-reversal (TR) symmetry. Experimental evidences of this unique gapless
one-dimensional ($1$D) quantum liquid are found based the the transport measurements.~\cite{Konig,Roth,CBrune}

\begin{figure}
\begin{center}
 \includegraphics[width=0.9\columnwidth]{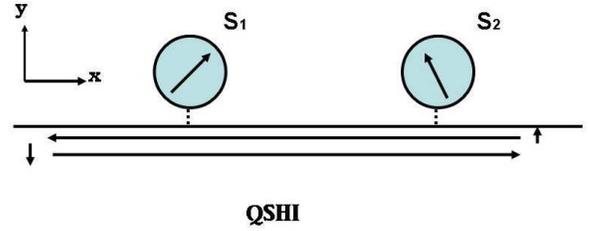}
 \caption{(Color online) Schematic setup of two local spins $\bm{S}_1$ and $\bm{S}_2$ located near the helical
 edge states of a QSHI which occupies the $xy$-plane with $y<0$. The local spins can be realized by quantum dots
 with strong on-site interactions. Both are exchange-coupled to the spin density of the helical edge states,
 denoted the dashed lines. The solid lines show the directions of the momenta of the edge states, with the
 associated spins indicated by the arrows.}
 \label{setup}
\end{center}
\end{figure}

In addition to the charge transport properties, the spin physics may provide an additional complementary signature
of the topological states of matter. This is particularly interesting for the edge states of the QSHI because the
spin polarizations are associated with the directions of their momenta. One way to detect this feature may be
provided by the Ruderman-Kittel-Kasuya-Yoshida (RKKY) interaction,~\cite{Kittel} which is an effective interaction
between two local spins mediated by itinerant electrons in the host metals. A good understanding on it may form
the basis for manipulating the quantum states of local spins, which can be crucial for the spintronics and quantum
computing.~\cite{GA,Simon} With this in mind, it is not surprising that the interplay between the RKKY interaction
and topological states of matter has attracted a lot of attentions. For examples, the RKKY interaction on the
surface of a three-dimensional topological insulator~\cite{QLiu,franz,JJZhu,pesin} and the edge of a QSHI~\cite{JGao}
have been studied. More recently, the RKKY interaction mediated by helical Majorana edge states in a TR-invariant
topological superconductor was also investigated.~\cite{EEriksson} In Ref. \onlinecite{JGao}, the RKKY interaction
mediated by the helical edge sate of the QSHI contains an in-plane non-collinear exchange coupling between two local
spins, in contrast to the isotropic coupling induced in normal metals.~\cite{Kittel}

It is known that the HgTe quantum wells in which the QSHI was first experimentally discovered exhibit some of the
largest known Rashba spin-orbit coupling (SOC) among semiconductor heterostructures.~\cite{Buh} Theoretically, it
was also shown that such a coupling in helical liquids leads to interesting new Kondo physics.~\cite{EEriksson2,Lee2}
Therefore, it is desirable to see how the Rashba coupling affects the RKKY interaction in the helical liquid.
Moreover, in the $1$D quantum liquid, it is well-known that the electron-electron interaction leads to the
breakdown of the Fermi liquid picture and the correct low-energy physics is described by Luttinger liquid (LL).~\cite{GNT}
Thus, a physically more accurate and realistic description of the RKKY interaction in the helical liquid must go
beyond the simplest model employed in Ref. \onlinecite{JGao} by taking into account both the above elements.

In this paper, we consider two local spins exchange-coupled to the spin density of the edge states of a QSHI. A
schematic setup is shown in Fig. \ref{setup}. We would like to study the effects of the Rashba SOC and the
electron-electron interaction on the RKKY interaction in the helical edge states. Since the strength of the Rashba
coupling can be tuned by an external gate voltage, we expect that the correlations between the two local spins may
be affected by the applied gate voltage through the Rashba SOC. Our main findings are as follows. (i) The spin
structure of the resulting RKKY interaction is fixed by the spin symmetries of the system. With the inclusion of
the Rashba coupling, it consists of the Heisenberg-like, the Dzyaloshinskii-Moria (DM), and the nematic-type terms.
The last one is absent by turning off the Rashba coupling. All terms contain the range functions of the forms
$\cos{(qx)}/|x|$ or $\sin{(qx)}/|x|$, where $|x|$ is the separation between the two local spins and $q=2\mu/\bar{v}_F$
with the chemical potential $\mu$ and the renormalized Fermi velocity $\bar{v}_F$. (ii) The electron-electron
interaction does not affect the spin structure of the RKKY interaction because the latter is determined by the spin
symmetries of the system. Its only effect is to modify the range functions which become $\gamma(K)\cos{(qx)}/|x|^{2K-1}$
or $\gamma(K)\sin{(qx)}/|x|^{2K-1}$, where $K$ is the LL parameter and the non-universal constant $\gamma(K)$
changes the sign from $K<1/2$ to $K>1/2$. (iii) By solving the two-impurity problem, we find that there exists a
critical strength of the electron-electron interaction corresponding to $K=1/2$. At that point a quantum phase
transition occurs because of the sign change of the range functions. (iv) From the solution of the two-impurity
problem, we may examine how the entanglement between the two local spins is affected by the Rashba coupling.
Since the latter can be tuned by the external gate voltage, our results provide a way to control the correlations
between the two local spins and may have further applications on spintronics and quantum computation.

The rest part of the paper is organized as follows. In Sec. \ref{model}, we setup our model and discuss the spin
symmetries of the system and the approximation we have made. The calculation of the RKKY by taking into account
the Rashba coupling and the electron-electron interaction is presented in Sec. \ref{rkky}. We solve the resulting
two-impurity problem in Sec. \ref{twoi}. Finally, the last section is devoted to a conclusive discussion.

\section{The model and the symmetries}
\label{model}

Our setup shown in Fig. \ref{setup} can be described by the Hamiltonian $H=H_0+H_{int}+H_{ex}$ when the energy
is much lower than the bulk gap, where $H_0$ is Hamiltonian for free edge electrons, $H_{int}$ gives the
interaction between them, whose form will be given later, and $H_{ex}$ describes the exchange interaction
between the local spins and the spin density of the helical edge states. By taking into account the Rashba SOC,
$H_0$ can be written as
\begin{eqnarray}
 H_0 &=& \! \int \! dx\Psi^{\dagger} \! \left[\bar{v}_F \! \left(\cos{\theta}\sigma_3i\partial_x+\sin{\theta}
 \sigma_2i\partial_x\right) \! -\mu\right] \! \Psi \nonumber \\
 &=& \! \int \! dx\Psi^{\dagger} \! \left(v_F\sigma_3i\partial_x+\alpha\sigma_2i\partial_x-\mu\right) \! \Psi \ ,
 \label{hlh1}
\end{eqnarray}
where $\Psi=[\psi_+,\psi_-]^t$ with $\psi_+=\psi_{L\uparrow}$ and $\psi_-=\psi_{R\downarrow}$, $v_F$ is the Fermi
velocity, $\alpha$ denotes the strength of the Rashba coupling, $\bar{v}_F=\sqrt{v_F^2+\alpha^2}$,
$\sin{\theta}=\alpha/\bar{v}_F$, and $\cos{\theta}=v_F/\bar{v}_F$. On account of the SOC present in this system,
$H_{ex}$ may be anisotropic, and thus we write it in the form
\begin{equation}
 H_{ex}=\! \sum_i \! \sum_{a=x,y,z} \! J_aS^a_i\hat{O}_s^a(x_i) \ , \label{rkkyh2}
\end{equation}
where $J_x=J_y=J_{\perp}\neq J_z$, $\hat{O}^a_s=\Psi^{\dagger}\sigma_a\Psi$ with $\sigma_{x,y,z}$ being Pauli
matrices, and $\bm{S}_i$ is the local spin at point $x_i$.

Before plunging into the calculation of the RKKY interaction, we investigate the spin symmetries of the
Hamiltonian $H_0+H_{ex}$. First of all, we notice that for $\theta\neq 0$ and $J_{\perp}=J_z$, $H_0+H_{ex}$ is
invariant against spin rotations about the $x$-axis, which will be dubbed as the spin U$_x$($1$) symmetry. This
can be shown as the following. Let $U_x(\phi)$ denote the spin rotation about the $x$-axis by angle $\phi$.
$S^x_j$ with $j=1,2$ and $\hat{O}_s^x$ are invariant, while $S_j^y$ and $S_j^z$ with $j=1,2$ transform like
\begin{eqnarray}
 U_x(\phi)S_j^yU_x^{\dagger}(\phi) &=& S_j^y\cos{\phi}+S_j^z\sin{\phi} \ , \nonumber \\
 U_x(\phi)S_j^zU_x^{\dagger}(\phi) &=& -S_j^y\sin{\phi}+S_j^z\cos{\phi} \ , \label{ux1}
\end{eqnarray}
and similar expressions for $\hat{O}_s^y$ and $\hat{O}_s^z$. On the other hand, $\Psi$ transforms like
\begin{equation}
 \tilde{\Psi}=U_x(\phi)\Psi=e^{-i\frac{\phi}{2}\sigma_x}\Psi \ . \label{ux11}
\end{equation}
In terms of Eqs. (\ref{ux1}) and (\ref{ux11}), one may show that
\begin{equation}
 U_x(\phi)[H_0(\theta)+H_{ex}]U_x^{\dagger}(\phi)=H_0(\theta-\phi)+H_{ex} \ , \label{ux12}
\end{equation}
if $J_{\perp}=J_z$.

Next, for $\theta=0$, the spin symmetry of $H_0+H_{ex}$ becomes the spin U$_z$($1$), i.e. the Hamiltonian
$H_0+H_{ex}$ is invariant against spin rotations about the $z$-axis. Finally, $H_0+H_{ex}$ has the exchange
symmetry. That is, it is invariant against the exchange of $\bm{S}_1$ and $\bm{S}_2$. We shall see later that
the three spin symmetries, the spin U$_x$($1$) symmetry, the spin U$_z$($1$) symmetry, and the exchange symmetry,
will impose constraints on the possible forms of the RKKY interaction.

\section{The RKKY interaction}
\label{rkky}

Now we are in a position to calculate the RKKY interaction in the helical liquid. We shall work in the
imaginary-time formulation. By integrating out the edge electrons, the resulting RKKY interaction takes the form
\begin{equation}
 H_{RKKY}=\sum_{a,b=x,y,z}J_aJ_b\Pi_{ab}(x)S_1^aS_2^b \ , \label{rkkih1}
\end{equation}
where $x=x_1-x_2$ and
\begin{eqnarray*}
 \Pi_{ab}(x)=-\! \int^{+\infty}_{-\infty} \! d\tau\langle\mathcal{T}_{\tau}\{\hat{O}_s^a(\tau,x_1)\hat{O}_s^a
 (0,x_2)\}\rangle \ ,
\end{eqnarray*}
is the Fourier transform of the spin-spin correlation function of the helical edge states at zero frequency and
zero temperature, with $\mathcal{T}_{\tau}$ denoting the time ordering in the imaginary-time formulation.

\subsection{The role of the Rashba coupling}

We first study the free helical liquid by ignoring $H_{int}$. For the free helical liquid, the spin-spin
correlation function is the product of two single-particle Green functions:
\begin{eqnarray*}
 \mathcal{S}_{12}^{ab}(\tau) \! \! &\equiv& \! \! -\langle\mathcal{T}_{\tau}\{\hat{O}_s^a(\tau,x_1)\hat{O}_s^a
 (0,x_2)\}\rangle \\
 \! \! &=& \! \! (\sigma_a)_{\alpha\beta}(\sigma_b)_{\lambda\rho}G_{\beta\lambda}(\tau;x_1,x_2)G_{\rho\alpha}
 (-\tau;x_1,x_2) \ ,
\end{eqnarray*}
where
\begin{eqnarray*}
 G_{\alpha\beta}(\tau;x_1,x_2)=-\langle\mathcal{T}_{\tau}\{\psi_{\alpha}(\tau,x_1)\psi^{\dagger}_{\beta}(0,x_2)\}
 \rangle \ ,
\end{eqnarray*}
is the single-particle Green function. With the help of the spectral representation of the single-particle Green
function
\begin{eqnarray*}
 \mathcal{G}_{\alpha\beta}(i\omega_n;x_1,x_2)=\! \int^{+\infty}_{-\infty} \! \frac{d\nu}{2\pi}
 \frac{\rho_{\alpha\beta}(\nu;x_1,x_2)}{i\omega_n-\nu} \ ,
\end{eqnarray*}
where $\mathcal{G}_{\alpha\beta}(i\omega_n;x_1,x_2)$ denotes the Fourier transform of
$G_{\alpha\beta}(\tau;x_1,x_2)$ with $\omega_n=(2n+1)\pi T$ and $\rho_{\alpha\beta}(\omega;x_1,x_2)$ is the
spectral function of electrons, one may write $\Pi_{ab}(x)$ as
\begin{eqnarray*}
 \Pi_{ab}(x)=W_{ab}(x)+W_{ba}(-x) \ ,
\end{eqnarray*}
at $T=0$, where
\begin{eqnarray*}
 W_{ab}(x) &=& \! \int^0_{-\infty} \! \frac{d\omega_1}{2\pi} \! \int^{+\infty}_{-\infty} \! \frac{d\omega_2}
 {2\pi}\frac{e^{\omega_10^+}}{\omega_1-\omega_2} \\
 & & \times\mbox{tr} \! \left\{\sigma_a[\rho](\omega_1;x)\sigma_b[\rho]
 (\omega_2;-x)\right\} .
\end{eqnarray*}
The rest of task is to determine the matrix $[\rho](\omega,x)$ for the helical liquid.

For the free helical liquid, we find that
\begin{eqnarray}
 [\rho](\omega;x) &=& -\frac{i}{\bar{v}^2_F}\sin{(\bar{\omega}x/\bar{v}_F)}(v_F\sigma_z+\alpha\sigma_y) \nonumber
 \\
 & & +\frac{1}{\bar{v}_F}\cos{(\bar{\omega}x/\bar{v}_F)}\sigma_0 \ , \label{sdos1}
\end{eqnarray}
where $\bar{\omega}=\omega+\mu$ and $\sigma_0$ is the $2\times 2$ unit matrix. In terms of Eq. (\ref{sdos1}), we
get
\begin{eqnarray}
 H_{RKKY} \! \! &=& \! \! V^{(1)}_{DM}(x)(\bm{S}_1\times\bm{S}_2)_z+V^{(2)}_{DM}(x)(\bm{S}_1\times\bm{S}_2)_y
 \nonumber \\
 \! \! & & \! \! +V_n(x) \! \left(S_1^yS_2^z+S_1^zS_2^y\right) \nonumber \\
 \! \! & & \! \! \! +\! \sum_{a=x,y,z} \! V_a(x)S_1^aS_2^a \ , \label{hlrkky1}
\end{eqnarray}
where the range functions are given by
\begin{eqnarray}
 V_x(x) &=& -\frac{J_{\perp}^2N(0)}{|x|}\cos{(qx)} \ , \nonumber \\
 V_y(x) &=& -\frac{J_{\perp}^2N(0)\cos^2{\theta}}{|x|}\cos{(qx)} \ , \nonumber \\
 V_z(x) &=& -\frac{J_z^2N(0)\sin^2{\theta}}{|x|}\cos{(qx)} \ , \nonumber \\
 V_n(x) &=& \frac{J_{\perp}J_zN(0)\sin{(2\theta)}}{2|x|}\cos{(qx)} \ , \nonumber \\
 V^{(1)}_{DM}(x) &=& \frac{J_{\perp}^2N(0)\cos{\theta}}{|x|}\sin{(qx)} \ , \nonumber \\
 V^{(2)}_{DM}(x) &=& \frac{J_{\perp}J_zN(0)\sin{\theta}}{|x|}\sin{(qx)} \ . \label{hlrkky11}
\end{eqnarray}
In the above, $q=2\mu/\bar{v}_F$ and $N(0)=1/(2\pi\bar{v}_F)$ is the density of states (DOS) at the Fermi level
for fermions with linear dispersion. The $V_x$, $V_y$, and $V_z$ terms are Heisenberg-like, the $V^{(1)}_{DM}$
and $V^{(2)}_{DM}$ terms are DM-like, and the $V_n$ term is the nematic type. The $1/|x|$ dependence of the range
functions is a characteristic of the $1$D free electrons.~\cite{Kittel}

A few comments on Eqs. (\ref{hlrkky1}) and (\ref{hlrkky11}) are in order. First of all, we notice that
$V_z(x)=0=V^{(2)}_{DM}(x)=V_n(x)$ in the absence of the Rashba coupling. This must be the case because all terms
in $H_{RKKY}$ arise from the backscattering of the host electrons. For the helical liquid, the backscattering
processes have to be accompanied by the spin flip of electrons. On the other hand, nonvanishing $V_z$,
$V^{(2)}_{DM}$, or $V_n$ terms imply the existence of backscattering processes which do not involve the spin flip
at least in one of the spin locations. When the Rashba SOC is turned on, the spin quantization axis is rotated so
that these terms are allowed. Next, we would like to show that the spin structure of $H_{RKKY}$ is fixed by the
spin symmetries of the system. Using Eq. (\ref{ux1}), one may verify that
\begin{eqnarray*}
 U_x(\phi)H_l(\theta)U_x^{\dagger}(\phi)=H_l(\theta-\phi) \ ,
\end{eqnarray*}
for $l=1,2$ when $J_{\perp}=J_z$, where
\begin{eqnarray*}
 H_1 &=& \! \sum_{a=y,z} \! V_a(x)S_1^aS_2^a+V_n(x) \! \left(S_1^yS_2^z+S_1^zS_2^y\right) , \\
 H_2 &=& V^{(1)}_{DM}(x)(\bm{S}_1\times\bm{S}_2)_z+V^{(2)}_{DM}(x)(\bm{S}_1\times\bm{S}_2)_y \ ,
\end{eqnarray*}
while the $V_x$ term is invariant against the spin U$_x$($1$) rotations. Hence, we conclude that
\begin{equation}
 U_x(\phi)H_{RKKY}(\theta)U_x^{\dagger}(\phi)=H_{RKKY}(\theta-\phi) \ , \label{ux13}
\end{equation}
when $J_{\perp}=J_z$. When $\theta=0$, it is straightforward to show that $H_{RKKY}$ respects the spin U$_z$($1$)
symmetry. Furthermore, $H_{RKKY}$ is invariant against the exchange of $\bm{S}_1$ and $\bm{S}_2$.

\subsection{The effects of the electron-electron interaction}

Now we take into account the electron-electron interaction. We shall assume that the helical liquid is still in
the LL phase in the presence of the electron-electron interaction. Thus, the most general form of $H_{int}$ is
given by
\begin{equation}
 H_{int}=\! \int \! \! dx \! \left(g_1 \! \sum_{\sigma=\pm}J_{\sigma}J_{\sigma}+g_2J_+J_-\right) , \label{hlh11}
\end{equation}
where $J_{\sigma}=\psi^{\dagger}_{\sigma}\psi_{\sigma}$. One may show that the rotated interacting Hamiltonian
$U_x(\phi)H_{int}U_x^{\dagger}(\phi)$ still takes the form of Eq. (\ref{hlh11}), except that the coupling
constant $g_1$ acquires the $\phi$ dependence. In this sense, the Hamiltonian $H=H_0+H_{int}+H_{ex}$ respects the
spin U$_x$($1$) symmetry. Consequently, Eq. (\ref{ux13}) still holds in the presence of $H_{int}$ when
$J_{\perp}=J_z$. Especially, we may take $\phi=\theta$ and it suffices to calculate $\Pi_{ab}(x)$ in the absence
of the Rashba SOC.

In terms of the bosonization formula~\cite{GNT}
\begin{eqnarray*}
 \tilde{\psi}_{\pm}(x)=\frac{1}{\sqrt{2\pi a_0}}e^{\mp i\mu x/\bar{v}_F}e^{\mp i\sqrt{4\pi}\phi_{\pm}(x)} \ ,
\end{eqnarray*}
where $a_0$ is the short-distance cutoff and $|\mu|/\bar{v}_F$ is the Fermi momentum, the helical liquid with
$\theta=0$ can be described by the Hamiltonian
\begin{equation}
 H=\frac{v}{2} \! \int \! \! dx\! \left[K(\partial_x\Theta)^2+\frac{1}{K}(\partial_x\Phi)^2\right] , \label{hlh2}
\end{equation}
at low energy, where $\Phi=\phi_++\phi_-$, $\Theta=\phi_+-\phi_-$, and $v$ is the speed of collective excitations.
$\Phi$ and $\Theta$ obey the commutation relation $[\Phi(x),\Theta(y)]=i\vartheta(y-x)$, where
$\vartheta(x)=1$, $1/2$, and $0$ for $x>0$, $x=0$, and $x<0$, respectively. The LL parameter $K=1$ in the absence
of $H_{int}$, while $K<1$ and $K>1$ for repulsions and attractions, respectively.

The actual value of $K$ depends on $\tilde{g}_1$ and $g_2$, where
\begin{eqnarray*}
 \tilde{g}_1=\! \left(1-\frac{1}{2}\sin^2{\theta}\right) \! g_1+\frac{1}{4}\sin^2{\theta}g_2 \ ,
\end{eqnarray*}
is the corresponding coupling constant in the transformed interacting Hamiltonian
$U_x(\theta)H_{int}U^{\dagger}(\theta)$. In the weak-coupling regime, we have
\begin{eqnarray*}
 K=\sqrt{\frac{1-\frac{g_2}{2\pi v_0}}{1+\frac{g_2}{2\pi v_0}}} \ , ~~
 v=v_0\sqrt{1- \! \left(\frac{g_2}{2\pi v_0}\right)^2} \ ,
\end{eqnarray*}
where $v_0=\bar{v}_F[1+\tilde{g}_1/(\pi\bar{v}_F)]$. Since $\tilde{g}_1$ depends on $\theta$, both $K$ and $v$
are functions of $\theta$ (and thus $\alpha$). Because the coupling to magnetic impurities may induce the
two-particle backscattering potential which will cut the helical edge state into two pieces and drive the system
into an insulating phase for $K<1/4$,~\cite{MLOQWZ,TFM} we shall restrict ourselves to the region with $K>1/4$
hereafter. Moreover, in order that the LL description is valid and the Kondo effect can be neglected, we must
require that $a_0\ll |x_1-x_2|\ll\bar{v}_F/T_K$, where $T_K$ is the Kondo temperature.

Upon bosonization, the components of the spin density operator $\hat{\bm{O}}_s$ can be written as
\begin{eqnarray}
 \hat{O}_s^x &=& \frac{i}{2\pi a_0}e^{2i\mu x/\bar{v}_F}e^{i\sqrt{4\pi}\Phi}+\mathrm{H.c.} \ , \nonumber \\
 \hat{O}_s^y &=& \frac{1}{2\pi a_0}e^{2i\mu x/\bar{v}_F}e^{i\sqrt{4\pi}\Phi}+\mathrm{H.c.} \ , \nonumber \\
 \hat{O}_s^z &=& \frac{1}{\sqrt{\pi}}\partial_x\Theta \ . \label{hlspin1}
\end{eqnarray}
In terms of Eq. (\ref{hlspin1}), the calculation of the spin-spin correlation functions is standard,~\cite{GNT}
and the nonvanishing components at zero temperature are
\begin{eqnarray}
 \mathcal{S}_{12}^{xx}(\tau) &=& -\frac{1}{2\pi^2a_0^2} \! \left[\frac{a_0^2}{(v|\tau|+a_0)^2+x^2}\right]^{\! K}
 \! \cos{(qx)} \nonumber \\
 &=& \mathcal{S}_{12}^{yy}(\tau) \ , \label{hlspin11} \\
 \mathcal{S}_{12}^{xy}(\tau) &=& \frac{1}{2\pi^2a_0^2} \! \left[\frac{a_0^2}{(v|\tau|+a_0)^2+x^2}\right]^{\! K}
 \! \sin{(qx)} \nonumber \\
 &=& -\mathcal{S}_{12}^{yx}(\tau) \ , \label{hlspin12}
\end{eqnarray}
and
\begin{equation}
 \mathcal{S}_{12}^{zz}(\tau)=-\frac{1}{2\pi^2K} \! \left\{\frac{(v|\tau|+a_0)^2-x^2}{[(v|\tau|+a_0)^2+x^2]^2}
 \right\} , \label{hlspin13}
\end{equation}
for $\theta=0$.

With the help of the above results and performing the spin U$_x$($1$) rotation, $H_{RKKY}$ still takes the form
of Eq. (\ref{hlrkky1}), but the range functions are replaced by the following ones:
\begin{eqnarray}
 V_x(x) &=& -\frac{J_{\perp}^2\gamma(K)N(0)}{|x|^{2K-1}}\cos{(qx)} \ , \nonumber \\
 V_y(x) &=& -\frac{J_{\perp}^2\gamma(K)N(0)\cos^2{\theta}}{|x|^{2K-1}}\cos{(qx)} \ , \nonumber \\
 V_z(x) &=& -\frac{J_z^2\gamma(K)N(0)\sin^2{\theta}}{|x|^{2K-1}}\cos{(qx)} \ , \nonumber \\
 V_n(x) &=& \frac{J_{\perp}J_z\gamma(K)N(0)\sin{(2\theta)}}{2|x|^{2K-1}}\cos{(qx)} \ , \nonumber \\
 V^{(1)}_{DM}(x) &=& \frac{J_{\perp}^2\gamma(K)N(0)\cos{\theta}}{|x|^{2K-1}}\sin{(qx)} \ , \nonumber \\
 V^{(2)}_{DM}(x) &=& \frac{J_{\perp}J_z\gamma(K)N(0)\sin{\theta}}{|x|^{2K-1}}\sin{(qx)} \ , \label{hlrkky2}
\end{eqnarray}
where $\gamma(K)=\frac{Ka_0^{2K-2}}{\sqrt{\pi}}\frac{\Gamma(K-1/2)}{\Gamma(K)}$ is a non-universal constant. We
see that these expressions reduce to those for the free helical liquid when $K=1$. The electron-electron
interactions reveal themselves in the $1/|x|^{2K-1}$ dependence and the prefactors of the range functions, as
expected for the $1$D interacting electrons.~\cite{ES,MDP} In contrast with the usual spin-$1/2$ LL and the
Rashba quantum wire, in the present case, the exponent of $1/|x|$ and the sign of $\gamma(K)$ may become negative
when $1/4<K<1/2$. We will see later that this sign change results in an impurity quantum phase transition for the
two-impurity problem.

\section{The two-impurity problem}
\label{twoi}

Here we consider the two-impurity problem. For simplicity, we consider the case with $J_{\perp}=J_z=J$ so that
$H_{RKKY}$ respects the spin U$_x$($1$) symmetry. To determine the ground state of $H_{RKKY}$, it suffices to
study the rotated Hamiltonian
\begin{eqnarray}
 & & \tilde{H}_{RKKY}\equiv U_x(\theta)H_{RKKY}U^{\dagger}_x(\theta) \label{twoih} \\
 & & =V(x) \! \left[\sin{(qx)}(\bm{S}_1\times\bm{S}_2)_z-\cos{(qx)} \! \! \sum_{a=x,y} \! S_1^aS_2^a\right] ,
     \nonumber
\end{eqnarray}
where $V(x)=J^2\gamma(K)N(0)/|x|^{2K-1}$.

\subsection{Classical spins}

We first consider the classical spins, which should be valid when $S\gg 1$. To proceed, we parametrize $\bm{S}_i$
as $\bm{S}_i=S(\sin{\theta_i}\cos{\phi_i},\sin{\theta_i}\sin{\phi_i},\cos{\theta_i})$ for $i=1,2$. Then, the
corresponding energy $E$ can be written as
\begin{eqnarray*}
 E(\theta_i,\phi_i)=-S^2V(x)\sin{\theta_1}\sin{\theta_2}\cos{(\phi_1-\phi_2-qx)} \ .
\end{eqnarray*}
For $K>1/2$, $V(x)>0$. In this situation the ground state is obtained when $\sin{\theta_i}=1$ for $i=1,2$ and
$\cos{(\phi_1-\phi_2-qx)}=1$, which corresponds to $\theta_i=\pi/2$ for $i=1,2$ and $\phi_1-\phi_2=qx$ module to
$2\pi$. On the other hand, for $1/4<K<1/2$, $V(x)<0$. In this situation the ground state is obtained when
$\sin{\theta_i}=1$ for $i=1,2$ and $\cos{(\phi_1-\phi_2-qx)}=-1$, which corresponds to $\theta_i=\pi/2$ for
$i=1,2$ and $\phi_1-\phi_2=qx+\pi$ module to $2\pi$. To sum up, the ground state in the rotated basis corresponds
to the planar spin structure $\bm{S}_i=S(\cos{\phi_i},\sin{\phi_i},0)$ for $i=1,2$ with the relative angle
$\phi_1-\phi_2=qx+n\pi$, where $n$ is some integer. By transforming to the original basis, the spin configuration
become
\begin{equation}
 \bm{S}_i=S(\cos{\phi_i},\sin{\phi_i}\cos{\theta},-\sin{\phi_i}\sin{\theta}) \ , \label{rkkys2}
\end{equation}
for $i=1,2$. We see that the spin configuration is not planar any more in the presence of the Rashba coupling.

Using Eq. (\ref{rkkys2}), we find that
\begin{eqnarray*}
 \langle\bm{S}_1\cdot\bm{S}_2\rangle=\cos{(\phi_1-\phi_2)}=\pm S^2\cos{(qx)} \ ,
\end{eqnarray*}
where the $+$ and $-$ signs correspond to $K>1/2$ and $1/4<K<1/2$, respectively, and $\langle\cdots\rangle$ means
the average over $\phi_2$ (or $\phi_1$). That is, the relative orientation between the two spins is independent
of the Rashba coupling. The latter reveals itself through other ``order parameters":
\begin{eqnarray*}
 \langle\bm{S}_1\times\bm{S}_2\rangle=\mp S^2\sin{(qx)}(0,\sin{\theta},\cos{\theta}) \ ,
\end{eqnarray*}
leading to the the chiral spin configuration, and
\begin{eqnarray*}
 Q_{ab} &=& \pm S^2\cos{(qx)} \! \left[\begin{array}{ccc}
 \frac{1}{3} & 0 & 0 \\
 0 & \cos^2{\theta}-\frac{2}{3} & -\frac{1}{2}\sin{(2\theta)} \\
 0 & -\frac{1}{2}\sin{(2\theta)} & \sin^2{\theta}-\frac{2}{3}
 \end{array}\right] ,
\end{eqnarray*}
leading to the nematic spin configuration, where
$Q_{ab}=\langle S_1^aS_2^b+S_1^bS_2^a-\frac{2}{3}\delta_{ab}\bm{S}_1\cdot\bm{S}_2\rangle$. In the above, the
upper and lower signs correspond to $K>1/2$ and $1/4<K<1/2$, respectively. Hence, we may tune the spin
configuration through the chemical potential and the strength of the Rashba coupling by external gate voltages.

\subsection{Spin-$1/2$}

Next, we turn into the spin-$1/2$ case. By choosing the basis as the eigenstates of the total spin
$\hat{\bm{S}}=\hat{\bm{S}}_1+\hat{\bm{S}}_2$, denoted by $|S,S_z\rangle$, the rotated Hamiltonian
$\tilde{H}_{RKKY}$ can be written as
\begin{eqnarray*}
 \tilde{H}_{RKKY}=\frac{V(x)}{2} \! \left[\begin{array}{cccc}
 \cos{(qx)} & i\sin{(qx)} & 0 & 0 \\
 -i\sin{(qx)} & -\cos{(qx)} & 0 & 0 \\
 0 & 0 & 0 & 0 \\
 0 & 0 & 0 & 0
 \end{array}\right] ,
\end{eqnarray*}
with the basis $\{|0,0\rangle,|1,0\rangle,|1,1\rangle,|1,-1\rangle\}$. It turns out that $|1,1\rangle$ and
$|1,-1\rangle$ are still the eigenstates of $\tilde{H}_{RKKY}$ with energies $E_{t,\pm}=0$. Nevertheless, the
states $|0,0\rangle$ and $|1,0\rangle$ are mixed. The resulting eigenstates are
\begin{eqnarray*}
 |0,+\rangle &=& i\cos{(qx/2)}|0,0\rangle +\sin{(qx/2)}|1,0\rangle \ , \\
 |0,-\rangle &=& -i\sin{(qx/2)}|0,0\rangle +\cos{(qx/2)}|1,0\rangle \ ,
\end{eqnarray*}
with energies $E_{0,\pm}=\pm V(x)/2$. We notice that all eigenstates of $\tilde{H}_{RKKY}$ are also the
eigenstates of $\hat{S}_z$ due to $[\hat{S}_z,\tilde{H}_{RKKY}]=0$. Since $V(x)>0$ for $K>1/2$ and $V(x)<0$ for
$1/4<K<1/2$, the ground state of $\tilde{H}_{RKKY}$ is $|0,-\rangle$ for $K>1/2$ and $|0,+\rangle$ for $1/4<K<1/2$.
Since the two states are orthogonal to one another, we expect that there exists an impurity quantum phase
transition by varying the value of $K$, which can be achieved by tuning the strength of the interaction or the
Rashba coupling.

In any case, the ground state has $S_z=0$ indicating the planar spin configuration, similar to the classical spins.
The ground state in the original basis is given by $|\Phi_{\pm}\rangle=U_x^{\dagger}(\theta)|0,\pm\rangle$. Using
$U^{\dagger}_x(\theta)|0,0\rangle=|0,0\rangle$ and the spin-$1$ representation of $U_x(\theta)$
\begin{eqnarray*}
 U_x(\theta)=\! \left(\begin{array}{ccc}
 \frac{\cos{\theta}+1}{2} & -\frac{i}{\sqrt{2}}\sin{\theta} & \frac{\cos{\theta}-1}{2} \\
 -\frac{i}{\sqrt{2}}\sin{\theta} & \cos{\theta} & -\frac{i}{\sqrt{2}}\sin{\theta} \\
 \frac{\cos{\theta}-1}{2} & -\frac{i}{\sqrt{2}}\sin{\theta} & \frac{\cos{\theta}+1}{2}
 \end{array}\right) ,
\end{eqnarray*}
we may write $|\Phi_{\pm}\rangle$ as
\begin{eqnarray}
 |\Phi_+\rangle &=& \frac{1}{\sqrt{2}}[i\cos{(qx/2)}+\cos{\theta}\sin{(qx/2)}]|+-\rangle \nonumber \\
 & & \frac{1}{\sqrt{2}}[i\cos{(qx/2)}-\cos{\theta}\sin{(qx/2)}]|-+\rangle \nonumber \\
 & & +\frac{i}{\sqrt{2}}\sin{\theta}\sin{(qx/2)}(|++\rangle+|--\rangle) \ , \label{twoig1} \\
 |\Phi_-\rangle &=& -\frac{1}{\sqrt{2}}[i\sin{(qx/2)}-\cos{\theta}\cos{(qx/2)}]|+-\rangle \nonumber \\
 & & -\frac{1}{\sqrt{2}}[i\sin{(qx/2)}+\cos{\theta}\cos{(qx/2)}]|-+\rangle \nonumber \\
 & & +\frac{i}{\sqrt{2}}\sin{\theta}\cos{(qx/2)}(|++\rangle+|--\rangle) \ . \label{twoig11}
\end{eqnarray}
In Eqs. (\ref{twoig1}) and (\ref{twoig11}), we have expanded $|\Phi_{\pm}\rangle$ by the states
$|S_1^z,S_2^z\rangle$ with $+$ and $-$ denoting spin-up and -down, respectively.

\begin{figure}
\begin{center}
 \includegraphics[width=0.9\columnwidth]{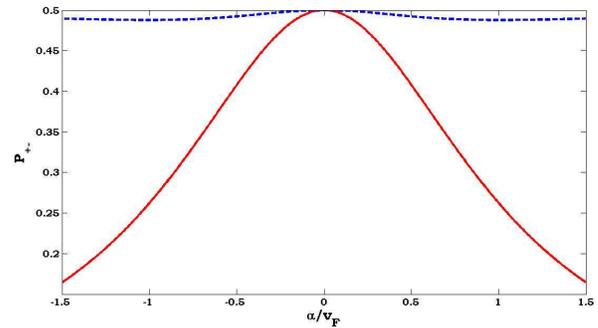}
 \caption{(Color online) The occupation probability $P_{+-}$ of the configuration $|+-\rangle$ as a function of
 $\alpha/v_F$. We have set $|\mu x|/v_F=0.1\pi$. The solid and dashed lines correspond to $K>1/2$ and $1/4<K<1/2$,
 respectively.}
 \label{p12}
\end{center}
\end{figure}

From Eqs. (\ref{twoig1}) and (\ref{twoig11}), we may tune the occupation probability of the two spins in some
configuration by the external gate voltages through the chemical potential or the Rashba coupling. For example,
the probability for the two spins in the configuration $|+-\rangle$ is given by
\begin{equation}
 P_{+-}=\frac{1}{2} \! \left[1-\sin^2{\theta}\cos^2{(qx/2)}\right] , \label{twoig12}
\end{equation}
for $K>1/2$ and
\begin{equation}
 P_{+-}=\frac{1}{2} \! \left[1-\sin^2{\theta}\sin^2{(qx/2)}\right] , \label{twoig13}
\end{equation}
for $1/4<K<1/2$. A plot of $P_{+-}$ as a function of $\alpha/v_F$ is shown in Fig. \ref{p12}. We have assumed that
the sign of $K-1/2$ does not change in the displayed values of $\alpha/v_F$. It is possible that $K-1/2$ will
change sign upon varying the value of $\alpha/v_F$. In that case, $P_{+-}$ will exhibit a discontinuous jump.

One may also calculate the expectation values of various ``order parameters" using $|\Phi_{\pm}\rangle$, yielding
\begin{eqnarray*}
 \langle\hat{\bm{S}}_1\cdot\hat{\bm{S}}_2\rangle &=& -\frac{1}{4}\pm\frac{1}{2}\cos{(qx)}  \ , \\
 \langle\hat{\bm{S}}_1\times\hat{\bm{S}}_2\rangle &=& \mp\frac{1}{2}\sin{(qx)}(0,\sin{\theta},\cos{\theta}) \ ,
\end{eqnarray*}
and
\begin{eqnarray*}
 \langle\hat{Q}_{ab}\rangle &=& \frac{1}{2} \! \left[\begin{array}{ccc}
 \frac{1}{3} & 0 & 0 \\
 0 & \cos^2{\theta}-\frac{2}{3} & -\frac{1}{2}\sin{(2\theta)} \\
 0 & -\frac{1}{2}\sin{(2\theta)} & \sin^2{\theta}-\frac{2}{3}
 \end{array}\right] \\
 & & \times [1\pm\cos{(qx)}] \ ,
\end{eqnarray*}
where the upper and lower signs correspond to $K>1/2$ and $1/4<K<1/2$, respectively. We see that the $\theta$
dependence of these ``order parameters" is identical to that for classical spins.

\section{Conclusions and discussions}
\label{con}

In summary, we have investigated the RKKY interaction between two local spins (quantum dots) mediated by the
helical edges state of a QSHI based on a model that includes both the Rashba SOC and the electron-electron
interaction. The former is an inevitable element in an asymmetric quantum well that realizes the QSHI, while the
latter is crucial for any $1$D electron system. Our results should be valid as long as the distance between the
local spins is longer than the one setup by the inverse of the bulk energy gap, while smaller than the size of
the Kondo screening cloud. Due to the breaking of the spin SU($2$) symmetry, the resulting exchange interaction
shows a nontrivial tensor structure, which is fixed by the spin symmetries of the system. Most importantly, by
electrically controlling the strength of the Rashba coupling, one may change the way how the two local spins
entangle together. Moreover, by varying the interaction strength, which may be achieved by tuning the strength
of the Rashba coupling, the excited and the ground states of the two spins may be switched, thus implying an
impurity quantum phase transition.

In comparison with the result obtained in Ref. \onlinecite{JGao} in which the Rashba SOC is not included, the
main effects of adding the Rashba coupling are to rotate the in-plane non-collinear exchange interactions about
the $x$-axis so that more non-collinear terms are generated, and the resulting range functions depend on the
strength of the Rashba coupling. In the absence of the Rashba coupling, the only factor that can be changed by
the external gate voltage is the chemical potential. By adding the Rashba interaction, we have more freedom to
control the correlations of local spins which should be important in spintronics and quantum computations.
Furthermore, these non-collinear terms may be used to engineer $1$D spin models which may exhibit non-trivial
spin orders due to the long range nature of the RKKY interaction and the presence of the SOC.~\cite{LPL}

For a non-interacting Rashba quantum wire, it has been established that the RKKY interaction becomes anisotropic,
and thus has a tensorial character.~\cite{IBU,Simonin,LLZ} Moreover, there are different spatial oscillation
periods reflecting the presence of different Fermi momenta in a Rashba quantum wire.~\cite{LLZ,egger} In the
present case, there is only one spatial oscillation period reflecting a unique Fermi momentum, which is distinct
from the usual Rashba quantum wire. Furthermore, practical calculations on the $1$D systems such as the Rashba
quantum wires,~\cite{egger} the carbon nanotubes,~\cite{KL} and the graphene nanoribbons,~\cite{KL} indicate
that not all tensor forms will appear in the RKKY interaction. Especially, the $V_n$ term of the nematic type is
unique for the helical liquid with the Rashba SOC. The reason is that the helical liquid itself already exhibits
a kind of SOC even in the absence of the Rashba coupling by breaking the spin SU($2$) symmetry down to the spin
U$_z$($1$) symmetry. The inclusion of the Rashba coupling changes the spin symmetry of the system and leads to
this nematic type interaction.

Finally, we notice that the RKKY interaction mediated by the helical Majorana edge states of a TR-invariant
topological superconductor was discussed in a recent work.~\cite{EEriksson} Due to the special feature of
Majorana fermions, the Fermi energy is always pinned at zero such that the spatial oscillation for the usual
RKKY interaction is absent. In that case, an impurity quantum phase transition was found due to the competition
between the exchange interaction mediated by the bulk gapped excitation of the topological superconductor and
that mediated by the gapless Majorana edge states. This is very different from our case where the quantum phase
transition is induced by varying strength of either the electron-electron interaction or the Rashba coupling.

\acknowledgments

The work of Y.-W. Lee is supported by the National Science Council of Taiwan under Grant No. NSC
99-2112-M-029-006-MY3.


\end{document}